\documentclass[report,floatfix,aps,nofootinbib]{revtex4}
\usepackage{epsfig}

\def\lp {\left( }
\def\rp {\right) }
\def\lb {\left[ }
\def\rb {\right] }

\def\nn {\nonumber}

\def\beq{\begin{equation}}
\def\eeq{\end{equation}}
\def\bea{\begin{eqnarray}}
\def\eea{\end{eqnarray}}
\def\ni{\noindent}

\def\m{\mu}
\def\n{\nu}

\begin{document}

\title{Quantum mechanics in curved space-time II}

\author{C.C. Barros Jr.}

\affiliation{
Instituto de F\'{\i}sica, Universidade de S\~{a}o Paulo,\\
C.P. 66318, 05315-970, S\~ao Paulo, SP, Brazil}

\date{\today}

\begin{abstract}

This paper is a sequence of the work presented in
 \cite{cb1}, where,
the 
 principles of the general 
relativity have been used to formulate quantum wave equations 
taking into account 
the effect of the electromagnetic and strong interactions in the
space-time metric of quantum systems.
Now, the role of the 
energy-momentum tensor in this theory is studied, and it is 
consistent with the formulation of the general quantum mechanics shown in 
\cite{cb1}. With this procedure, a dynamical cut-off is generated
and the constant $A$ of the field equation is calculated.
\end{abstract}

\maketitle

\vspace{5mm}

\section{introduction}

A major question in the formulation of physics is if the concepts used to 
study  large scale systems, as galaxies and even in the cosmology, may be
 applied to very small systems, such as atoms and elementary particles.
 Should 
the basic principles of physics depend on the size of the  object?

In practical terms, one aspect of 
this question is if the microscopic world
interactions, the weak, electromagnetic and the strong, may affect 
significantly the space-time metric and if this proposition may have any
observable effect. In this description, the gravitational forces may be 
neglected, and it is a good approximation,
due to the small masses of the considered particles.
 In \cite{cb1}, these ideas 
have been formulated  
considering a particle in a region with a potential, that affects the metric,
 and the wave equations for spin-0 and
spin-1/2 particles, based on the general relativity principles,
 have been proposed,  generating very interesting results. 

The simplest systems where this theory could be tested  are the one 
electron atoms, and the calculation of the deuterium spectrum has shown
a clear numerical improvement when compared with the usual
Dirac spectrum \cite{dir1}, \cite{scf} (also proposed by Sommerfeld 
\cite{som}),  with a percentual
 deviation from the experimental results 
approximately five times 
smaller, near
  one additional digit of precision.

An interesting fact that appeared from this theory, is the existence of black
holes inside these quantum systems, 
 with sizes that are not negligible. This propriety, that is related with 
the existence of a
trapping surface at $r_0$, as it was defined by Penrose \cite{pen},
 in \cite{cb2}
have been successfully used in order to describe quark confinement.
Solving the quantum wave equations \cite{cb2},  quark confinement
has been obtained, without the need of introducing confining potentials, as it 
is currently done \cite{bog}- \cite{zz}. The confinement obtained in this way 
is a strong confinement,  
as the quark wave functions are not continuos at $r_0$.

In order to compliment this theory, proposed in \cite{cb1} and \cite{cb2},
 the energy-momentum tensor must be included
in this formulation, and this study is the 
main purpose of this paper. We look for consistence with the previous results 
in an approach based on the energy-momentum tensor $T^{\m\n}$.

\section{Field equations and metric}

In this section, the field equations for particles subjected to 
electromagnetic and strong interactions, neglecting the effect of the
gravitational field, will be obtained. For this purpose we will 
 make a brief review of the results of \cite{cb1}, and  then, relate them with
a formulation based on the energy-momentum tensor.

As a first step,  a system with spherical symmetry will be considered,
 but the 
basic ideas can be generalized to systems with arbitrary metrics. 
If the spherical symmetry is considered, 
 the space-time may be described 
by a Schwarzschild-like metric \cite{lan},\cite{wein},
\beq
ds^2=\xi\ d\tau^2 - r^2(d\theta^2+ \sin^2 \theta\ d\phi^2) - \xi^{-1}dr^2  \ 
 \  ,
\label{I.1}
\eeq 

\ni
where $\xi(r)$ is determined by the interaction potential $V(r)$, and is a 
function only of 
$r$, for a time independent interaction. 
So, the metric tensor $g_{\m\n}$ is diagonal 
\beq
g_{\m\n}=\pmatrix{\xi&0&0&0\cr 0&-\xi^{-1}&0&0\cr  0&0&-r^2&0\cr
0&0&0&-r^2\sin^2\theta\cr}  \  \  .
\label{gmn}
\eeq

 The energy relation for this system is \cite{cb1} 
\beq
{E\over \sqrt{\xi}}=\sqrt{p^2 c^2 + m_0^2 c^4}  \  \ ,
\label{en}
\eeq

\ni
therefore,
\beq
E(\vec\beta=0)=E_0\xi^{1/2}=E_0+V  \  \  ,
\eeq

\ni
where
$\vec\beta$ is the particle velocity. This relation means  that
in the rest frame of the particle, the energy is simply due to 
the sum of its 
rest mass $E_0$ with  the potential, and then, 
\beq
\xi^{1\over 2} = 1+{V\over mc^2}  \  \  .
\label{ximet}
\eeq

\ni
Applying these ideas in the study of one electron atoms, one must
consider a Coulomb interaction
\beq
V(r)=-{\alpha \ Z\over r}  \  ,
\eeq

\ni
and $\xi$ is given by
\beq
\xi =1 -{2\alpha \ Z\over mc^2r}+{\alpha^2Z^2\over m^2c^4r^2} \  .
\label{xi1}
\eeq

\ni
These expressions determine
 the horizon of events at $r_0$, that appears from the
metric singularity $\xi(r_0)$=0, and using the values of \cite{pdg},
one finds
\beq
r_0= {\alpha Z\over mc^2}=2.818\ Z \  {\rm fm}  \  \  ,
\label{r0}
\eeq

\ni that is not a negligible value at the atomic scale.

Now, let us turn our attention to a description based on the energy-momentum
tensor. 
If one consider a field generated by
  the electromagnetic interaction, the energy-momentum tensor is
\beq
T^{\m\n}= \epsilon_0 \lp F^{\m\alpha} F_{\alpha}^\n -{1\over 4} 
\delta^{\m\n} F_{\alpha\beta}F^{\alpha\beta}  \rp  \   ,
\eeq

\ni
and it is related to the space-time geometry by 
the field equations
\beq
R_\n^\m-{1\over 2}R\delta_\n^\m=-AT_\n^\m  \  ,
\label{curv}
\eeq

\ni
where $A$ is a constant to be determined.

\ni
In an one electron atom, 
$T^{\m\n}$ is determined by an electrostatic field, 
with the nonvanishing components
\beq
T_0^0= {1\over 2} \lp \epsilon_0E^2+{B^2\over \m_0}   \rp  \   , 
\eeq

\ni
and $T_r^r=-T_0^0$,
that for a central electrostatic field with charge $Ze$, $T_0^0$ 
is just
\beq
T_0^0={\epsilon_0E^2\over 2}= {Z^2\alpha\over 8\pi r^4}  \  \  .
\label{t00}
\eeq

Eq. (\ref{curv}) may be used to determine the $\xi$ function
observing that in the given metric
\bea
&&{e^{-B}\over r^2} \lp r{dB\over dr}-1   \rp +{1\over r^2}=AT_0^0  \nn  \\
&&{e^{-B}\over r^2} \lp r{dB\over dr}+1   \rp -{1\over r^2}=-AT_0^0  \  \  ,
\label{ein1}
\eea

\ni
that have the solution \cite{mmal}
\beq
\xi(r)=e^{-B}=1-{c^2AM(r)\over 4\pi r}
\eeq

\ni
with
\beq
M(r)=\int_0^r{4\pi\over c^2} (r')^2 T_0^0 dr'   \   .
\label{mrr}
\eeq

\ni
If the particle is outside the black hole, that is the case of the electron
in an atom, it will be affected just by the part of the field located
in the region  external  to the horizon of events, and the integration 
(\ref{mrr}) must be performed in the interval $r_0\leq r'\leq r$,
\beq
M(r)=m_0+\int_{r_0}^r {4\pi\over c^2} \ (r')^2 T_0^0(r') dr'=
m_0+{Z^2\alpha\over 2\ c^2}
\lp{1\over r_0}-{1\over r}\rp  \  ,
\label{mmm}
\eeq
\ni
where $m_0$ is a constant of integration. So,
\bea
&& M(r) = m_0 + {mZ\over 2} -{Z^2\alpha\over 2\ c^2r}    \\
&& M(\infty)=m_0+ {mZ\over 2} 
\label{mm}
\eea

\ni
and
\beq
\xi= 1-{c^2AZ\over 4\pi r}\lb m_0+ {mZ\over 2} \rb + 
{AZ^2\alpha\over 8\pi r^2}
\label{mr}
\eeq 

\ni
that is solution of (\ref{ein1}).
The constants $A$ and $m_0$ may be obtained now,  comparing the expression 
(\ref{mr}), that is determined by the energy-momentum tensor and the
field equations, with (\ref{xi1}), determined by the energy relation 
(\ref{en}).

Identifying the terms $r^{-1}$ and $r^{-2}$ we have
\bea
A&=&{8\pi\alpha\over m^2c^4}   \nn \\
m_0&=&{mZ\over 2}   \   .
\eea

\ni
The field equation is then
\beq
R_\n^\m-{1\over 2}R\delta_\n^\m=-{8\pi\alpha\over m^2c^2}T_\n^\m  \  .
\label{bfld}
\eeq

So, in this system, due to the interaction,
the electron energy decrease with $r$, until it reaches the 
value $E(r_0)=0$ at the horizon of events. At large distances, $E(\infty)=m$.

Considering that the strong interactions may be approximated by a
strong coulomb field \cite{cb2}, with $\alpha\sim 1$, one may use 
the field equation
 (\ref{bfld}) in order to study strongly interacting systems.
 But if one wants to
 describe the strong interactions
by the Yang-Mills field, (\ref{bfld}) is determined by the correct
coupling constant $\alpha$, and an Yang-Mills energy-momentum tensor.

\section{Summary and conclusions}

In this paper the study of quantum systems in curved spaces 
has been continued.
The metric is determined by the interaction of quantum objects, such as
electrons and quarks. The effect of the
energy-momentum tensor of electromagnetic and strong
interactions in the space-time has been considered, and 
with this procedure the 
the constant of the field equations has been calculated.
The results obtained are consistent with the ones of \cite{cb1}.

One can observe the presence of the mass in the constant $A$, what  does
not happen in the general relativity, fact that is due to the equivalence 
principle. Observing that
 $A\propto g/m^2$, one concludes that the effect of the curvature
of space-time decreases for large masses and increases for small masses
and for large coupling constants. It is interesting to note that 
a dynamical cut-off is determined by this theory, as it was used in  eq. 
(\ref{mmm}), providing correct results.

Another observation that must be made is that
initially, spherical symmetry has been supposed, but eq. (\ref{bfld})
 is general, independent of the symmetry of the system.
This equation may also be used with the inclusion of other interactions,
as the Yang-Mills one, for example, considering the 
Yang-Mills field tensor $F_{\m\n}^a$ in the construction of the energy-momentum
tensor, and quark confinement, from the results of this theory,
is expected to  occur.

One must remark that with the development of the general quantum
mechanics, we are being able 
to explain many characteristics of the studied physical systems, 
using the new proprieties that appears from the understanding
of the geometry of the space-time.

\begin{acknowledgments}
I wish to thank  professor M. R. Robilotta.
This work was supported by FAPESP.
\end{acknowledgments}




\end{document}